\documentclass{Interspeech}

\interspeechcameraready

\title{In This Environment, As That Speaker: A Text-Driven Framework for Multi-Attribute Speech Conversion}

\author[affiliation={1}]{Jiawei}{Jin}
\author[affiliation={1}]{Zhihan}{Yang}
\author[affiliation={1}]{Yixuan}{Zhou}
\author[affiliation={1,\ast}]{Zhiyong}{Wu}

\affiliation{Shenzhen International Graduate School}{Tsinghua University}{China}
\email{jjw23@mails.tsinghua.edu.cn,zywu@sz.tsinghua.edu.cn \thanks{$\dagger$ Corresponding author.}}

\usepackage{comment}
\usepackage{hyperref}

\begin{document}

\maketitle
\renewcommand{\thefootnote}{\fnsymbol{footnote}}
\footnotetext[1]{Corresponding author.}
\renewcommand{\thefootnote}{\arabic{footnote}}

\begin{abstract}
We propose TES-VC (Text-driven Environment and Speaker controllable Voice Conversion), a text-driven voice conversion framework with independent control of speaker timbre and environmental acoustics. TES-VC processes simultaneous text inputs for target voice and environment, accurately generating speech matching described timbre/environment while preserving source content. Trained on synthetic data with decoupled vocal/environment features via latent diffusion modeling, our method eliminates interference between attributes. The Retrieval-Based Timbre Control (RBTC) module enables precise manipulation using abstract descriptions without paired data. Experiments confirm TES-VC effectively generates contextually appropriate speech in both timbre and environment with high content retention and superior controllability which demonstrates its potential for widespread applications. Our demo is avaliable at \url{https://thuhcsi.github.io/TES-VC/}.
\end{abstract}

\section{Introduction}
Voice Conversion (VC), which modifies speaker characteristics while preserving linguistic content, is undergoing a paradigm shift driven by advanced digital content needs.
Emerging applications are increasingly demanding simultaneous control over both vocal attributes and environmental acoustics, such as film dubbing \cite{zhang2024speaker,liu2024diffdub}, immersive AR/VR experiences \cite{ahmed2024voick}, and personalized voice interfaces \cite{kurniawati2012personalized}.
When dubbing a film, dynamic voice timbre adaptation is required to reflect character traits combined with acoustic environment modification to synchronize with narrative scenarios, collectively enhancing expressive dimensionality. 
This multi-dimension control challenge fundamentally redefines the design objectives for next-generation VC frameworks.

Text-driven VC has emerged as a more flexible and user-friendly solution to these demands, overcoming the constraints of reference speech-based \cite{park2023triaan,wu2020vqvc+,wu2020one,wang2021vqmivc,zhang2023leveraging,li2023styletts,hussain2023ace} and categorical label-based 
\cite{zhou2020converting,liu2019unsupervised} methods. Text-driven VC achieves speech style transformation by establishing cross-modal mappings between natural language descriptions and acoustic characteristics, converting source speech into target vocal styles that precisely align with textual specifications. HybridVC \cite{niu2024hybridvc} implements hybrid control of text and reference speech through contrastive learning mechanisms, enabling coherent guidance from both modalities. InstructVC \cite{kuan2023towards} enhances semantic comprehension of complex textual descriptions by integrating large language models (LLMs). PromptVC \cite{yao2024promptvc} employs a latent diffusion architecture to generate style vectors conditioned on text prompts, achieving multi-attribute control over timbre, intensity, pitch, and speaking rate while resolving the one-to-many mapping challenge through probabilistic sampling.
All these advancements allow systematic vocal customization for contextual narratives or personal preferences, significantly progressing personalized speech synthesis. 

Nevertheless, existing text-driven VC systems exhibit a critical limitation: excessive focus on intrinsic speech attributes (e.g., timbre, emotion) with inadequate attention to extrinsic environmental acoustics modeling (e.g., ``echoic cathedrals", ``crowded cafés"). Although recent attempts through room impulse response extraction \cite{tan2021environment} or acoustic mask construction~\cite{lu2024incremental} aim to expand environmental control, these methods primarily duplicate static features like steady-state noise while failing to model dynamic soundscape interactions (e.g., transient car sound, crowd murmurs). Notably, while VoiceLDM~\cite{lee2024voiceldm} achieves text-driven speech generation with environmental effects, its VC adaptation encounters CLAP-based limitations \cite{elizalde2023clap}: the cross-modal alignment mechanisms experience severe feature suppression \cite{zhang2024learning} when processing multi-event inputs, often discarding fine-grained features, especially the speaker characteristics.

To address these gaps, we extend text-driven VC for simultaneous environmental and vocal control, which confronting two key challenges. Firstly, current datasets lack parallel annotations for both speaker characteristics and environmental descriptors, with existing prompt-speech dataset’s \cite{guo2023prompttts,kawamura2024libritts} timbre-related labels (e.g., gender, age) being too coarse to support abstract descriptors like ``deep" or ``hoarse". Secondly, real-world audio recordings exhibit tightly entangled vocal-environment coupling, necessitating conversion mechanisms that (i) strictly preserve linguistic content integrity, (ii) achieve mutually orthogonal control over timbre and environmental acoustics through decoupled feature manipulation, and (iii) synthesize natural waveforms maintaining naturalness and realism.


To tackle all these problems, we proposed TES-VC, a text-driven voice conversion model with both timbre and environment control ability, with the following main contributions:
\begin{figure*}[t] 
    \centering
    \includegraphics[scale=0.5]{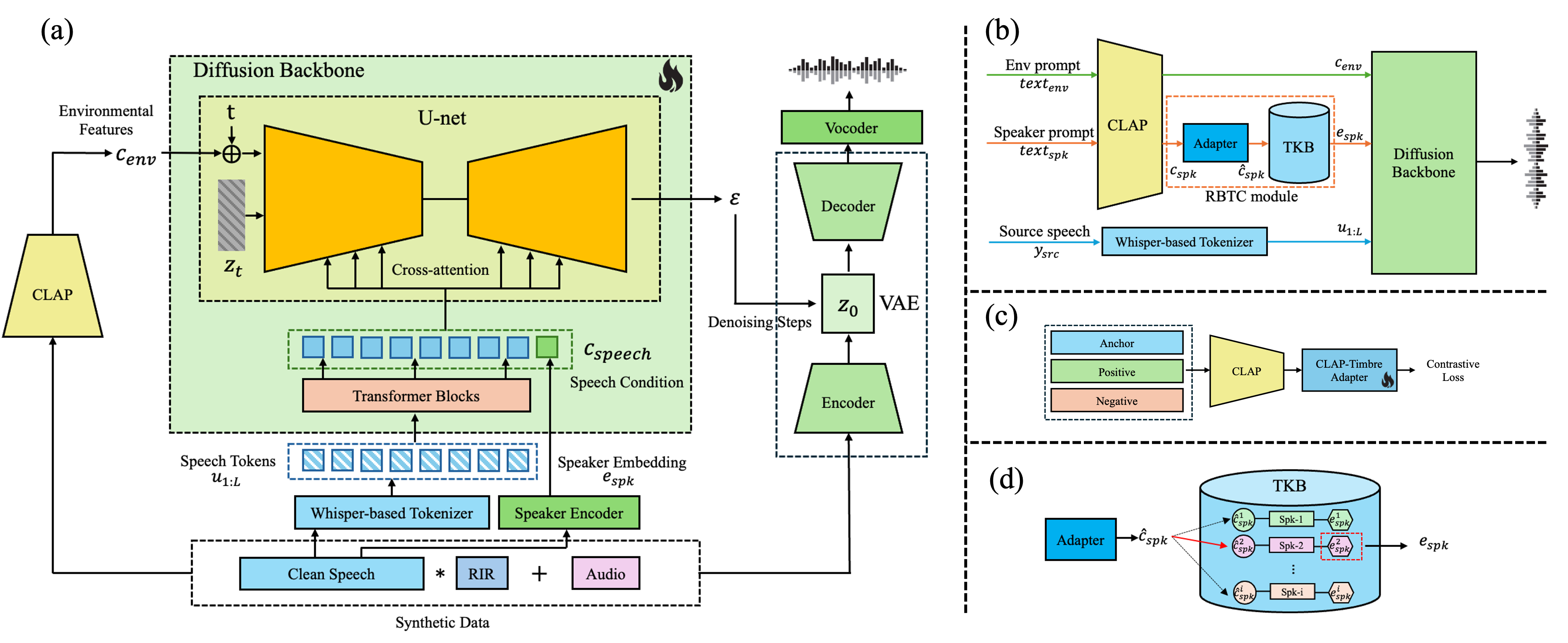} 
    \caption{The details of the architecture training stages and inference stages of our proposed method. Subfigure (a) is the architecture and the training stage of the diffusion backbone of TES-VC. Subfigure (b) is the inference stage of the proposed TES-VC. Subfigure (c) is the training stage for the proposed CLAP-Timbre Adapter. Subfigure (d) is the details of the proposed RBTC module. The flame icon represents the portion that the parameter updating is activated in the training.}
    \label{fig:example}
\end{figure*}
\begin{itemize}[leftmargin=2.2em, labelwidth=1.8em, labelsep=0.4em, nosep]
    \item[(1)] We establish a text-driven VC framework enabling simultaneous control of environmental acoustics and vocal timbre through natural language descriptions, meeting the high-level practical demands.
    \item[(2)] We implement disentangled modeling by generating synthetic data to circumvent the inherent vocal-environment coupling in real recordings, combined with a latent diffusion architecture that achieves independent encoding of environmental/timbral/content features to eliminate cross-dimensional interference.
    \item[(3)] We introduce a Retrieval-Based Timbre Control (RBTC) module that realizes fine-grained text-to-timbre mapping at the inference stage, which enables precise control of timbre and decreases the reliance on textual annotations.
\end{itemize}

\section{Proposed Method}
\subsection{System Overview}
The proposed TES-VC model is trained on purely acoustic data (Figure 1(a)), and leverages text-guided control during inference (Figure 1(b)).  As depicted in Figure 1(b), our model consists of a diffusion backbone and a series of feature extraction modules. The diffusion backbone generates transformed audio by processing three inputs: environmental features $c_{env}$ from text prompts $text_{env}$, speaker embeddings $e_{spk}$ from speaker prompts $text_{spk}$, and content tokens $u_{1:L}$ encoding source speech $y_{src}$. This framework integrates the CLAP model for cross-modal alignment with a Whisper-based speech tokenizer for content analysis. Our custom Retrieval-Based Timbre Control (RBTC) module enhances timbre manipulation through a CLAP-timbre adapter and Timbre Knowledge Base (TKB). All these modules operate in conjunction with each other where necessary to accurately extract essential information from the input, while discarding redundant data.

\subsection{Data Construction and Feature Extraction}
In the task of controlling both the environment and speaker, it is crucial to accurately decouple three key factors: timbre, environmental conditions, and speech content. However, this decoupling is challenging when training on real-world data, as these factors are often intertwined in natural recordings. To address this, we propose a data-feature strategy (Figure 1(a)) that selectively isolates features while generating realistic construction audio. Inspired by work in speech enhancement \cite{ephraim2006recent}, we model real-world speech as $y_{real}=y_{clean}*RIR+y_{audio}$, where $y_{clean}$ is the clean speech, $RIR$ is the room impulse response, and $y_{audio}$ represents non-speech audio events. Using this framework, we employ CLAP to encode synthetic audio to obtain an environmental embedding $c_{env}$, which captures both non-speech audio events and the interaction between speech and the environmental sound field. Our pipeline processes clean speech through a 12.5Hz Whisper-based tokenizer \cite{zeng2024glm} and DNN-based speaker encoder \cite{snyder2018x}, extracting content-preserving speech tokens $u_{1:L}$ and speaker-discriminative embeddings $e_{\text{spk}}$.

\subsection{Diffusion Backbone}
As illustrated in Figure 1(a), our diffusion backbone employs a U-Net architecture to iteratively denoise Gaussian-corrupted latent speech representations (encoded by pre-trained VAE) toward the target output. At each timestep $t$, the network predicts the noise component conditioned on three disentangled inputs: environmental embeddings $c_{env}$, speaker embeddings $e_{spk}$, and speech tokens $u_{1:L}$.
To align the highly compressed discrete content tokens with the continuous latent space, we first process them through a four-layer Transformer enhanced with rotary positional encoding, which generates contextualized hidden states $\tilde{u}_{1:L}$. Subsequently, these hidden states are concatenated with speaker embeddings $e_{spk}$ to form a composite conditioning signal $c_{speech}=cat(\tilde{u}_{1:L},e_{spk})$ which is then integrated into the denoising U-Net via cross-attention mechanisms. This architecture enables synchronized control over both content preservation and style transformation.
The training objective minimizes the re-weighted loss between predicted and ground-truth noise at each step:
\begin{align}
  L_\theta=||\epsilon-\epsilon_\theta(z_t,c_{env},c_{speech})||_2^2
\end{align}
As shown in Figure 1(b), during inference, the diffusion backbone takes three conditions $c_{env}$, $e_{spk}$, $u_{1:L}$ and performs denoising steps similar to training. To better balance speech and sound effects in the generated audio while enabling more flexible and diverse control, dual classifier-free guidance is employed, and the diffusion score is adjusted as:
\begin{align}
\tilde{\epsilon}_\theta&(z_t,c_{env},c_{speech})=\epsilon_\theta(z_t,c_{env},c_{speech})\\\nonumber
&+\omega_{env}(\epsilon_\theta(z_t,c_{env},\emptyset_{speech})-\epsilon_\theta(z_t,\emptyset_{env},\emptyset_{speech}))\\
&+\omega_{speech}(\epsilon_\theta(z_t,\emptyset_{env},c_{speech})-\epsilon_\theta(z_t,\emptyset_{env},\emptyset_{speech}))\nonumber
\end{align}
where $\emptyset$ is the null condition and $\omega_{env}$ and $\omega_{speech}$ are the guidance scale for environment and speech conditions, respectively.

\subsection{Retrieval-based Timbre Control Module}
The CLAP model’s robust audio-text alignment capabilities enable a data-efficient training paradigm where environmental control is learned directly from audio data (Figure 1(a)), while still allowing flexible text-guided control during inference (Figure 1(b)). This approach significantly reduces dependency on annotated text-audio pairs while improving controllability and generalization. However, CLAP’s event-level semantic representations inherently entangle fine-grained speaker characteristics with environmental and other acoustic features in its embedding space, making direct adaptation for speaker timbre control highly challenging. 
While there exits an one-to-one correspondence between CLAP’s speech representation $c_{spk}$ and the speaker’s timbre feature $e_{spk}$ for each speech, direct mapping via regression is difficult as CLAP's representation inherently loses fine-grained acoustic information during its semantic alignment process. 
Inspired by recent retrieval-augmented generation (RAG) strategies, we propose a retrieval-based timbre control module (RBTC) that bridges $c_{spk}$ and $e_{spk}$ which can further use CLAP’s cross-modal alignment to map from $text_{spk}$ to $e_{spk}$.

To extract speaker-specific representations, we introduce an adapter comprising an MLP and a Transformer-based fusion module. This adapter filters out speaker-irrelevant information from CLAP embeddings and transforms their distribution to better align with timbre control objectives. As shown in Figure 1(c), speech from the same speaker is used as positive samples and from different speakers as negative samples. The adapter is optimized using contrastive loss:
\begin{align}
    L_{contra}=max(SIM(\tilde{c}_a,\tilde{c}_p)-SIM(\tilde{c}_a,\tilde{c}_n)+\alpha,0)
\end{align}
where, $\tilde{c}_a$, $\tilde{c}_p$ and $\tilde{c}_n$ are the adapted CLAP embedding of the anchor, positive sample and negative sample, while $SIM(\cdot)$ refers to the cosine similarity.

On this basis, as shown in Figure 1(d), by constructing a CLAP-timbre knowledge base containing sample pairs $(\tilde{c}_{spk}^i,spk_i,e_{spk}^{i})$ at the speaker level, the optimal speaker embedding can be obtained through correlation-based retrieval:
\begin{align}
    e_{spk}^i = \arg\max_{i} \left( \frac{\tilde{c}_{spk} \cdot \tilde{c}_{spk}^i}{||\tilde{c}_{spk}||\cdot||\tilde{c}_{spk}^i||} \right)
\end{align}
where, $\tilde{c}_{spk}$ is the input query and $\tilde{c}_{spk}^i$ is the adapted CLAP embedding for the $i-th$ speaker in TKB. During inference, the input speaker description is encoded into the speech space through CLAP's cross-modal alignment, then RBTC module retrieves the speaker whose timbre most closely matches the description, and the corresponding $e_{spk}$ is used to achieve timbre control. 

\section{Experiments}
\subsection{Dataset}
To systematically decouple environmental, content, and timbral features while ensuring realistic audio synthesis, we implement the dynamic data construction strategy described in Section 2.2. Our training corpus comprises three components: (1) speech data from the LibriTTS-R train-clean-360 subset \cite{koizumi2023libritts}; (2) environmental audio from a filtered AudioSet subset (AS-audio \cite{lee2024voiceldm}); (3) pre-generated room impulse responses (RIRs) synthesized with GPU-RIR \cite{diaz2021gpurir}. All audio is resampled to 16 kHz and segmented into 10-second clips. During training, synthetic mixtures are generated on-the-fly with randomized augmentation: RIR convolution and environmental sound mixing (0.5 probability for each), silent speech inputs (0.2 probability). Meanwhile, to ensure sufficient speaker diversity, we utilize the entire LibriSpeech-train dataset, comprising over 2,000 speakers, for training the CLAP-timbre adapter and constructing the timbre knowledge base.
\subsection{Configuration}
We follow the setting of VoiceLDM for the U-Net backbone and train the proposed model from scratch. We train the diffusion backbone with a batch size of 4 for each using 8 NVIDIA H800 GPUs. The diffuison backbone are trained for 500K steps with a fixed learning rate at $2e\text{-}5$ for the AdamW optimizer. The CLAP-timbre adapter is trained with a batch size of 8 for 50k steps. During inference, a DDIM sampler with total inference steps of 50 are used for the denoising process.
\begin{table*}[t]
  \caption{Experimental results of subjective evaluation for different methods. The results of mean opinion scores (MOS) in the user study are shown with 0.95 confidence intervals. The last row shows the ablation study and “w/o CLAP-timbre adapter” refers to the model without the CLAP-timbre adapter in the RBTC module.}
  \label{tab:example}
  \centering
    \begin{tabular}{c c c c c c}
    \toprule
    \textbf{Systems} & \textbf{Acc\textsubscript{Gen}$\uparrow$} & \textbf{Acc\textsubscript{Age}$\uparrow$} & \textbf{S-MOS$\uparrow$} & \textbf{E-MOS$\uparrow$} & \textbf{R-MOS$\uparrow$} \\
    \midrule
    VoiceLDM\cite{lee2024voiceldm}        & 0.78  & 0.46     & $2.18 \pm 0.14$     & $3.01 \pm 0.22$ & $1.01 \pm 0.13$  \\
    \textbf{TES-VC}            & \textbf{0.88}  & \textbf{0.62}     &  $\bm{3.63 \pm 0.15}$      & $\bm{3.50 \pm 0.17}$ & $\bm{3.17 \pm 0.14}$  \\
    \midrule
    w/o CLAP-timbre adapter   & 0.84  & 0.58     & $3.17 \pm 0.18$      & $3.48 \pm 0.16$ & $3.11 \pm 0.09$  \\
    \bottomrule
    \end{tabular}
\end{table*}
\subsection{Experiment Results}
To the best of our knowledge, TES-VC is the first voice conversion framework capable of text-guided manipulation of both non-vocal environmental sounds and speaker timbre without requiring text-speech paired data. Given the scarcity of both task-aligned datasets and prior models achieving such integrated functionality, we meticulously design subjective and objective evaluations to assess three critical capabilities: precision in speaker timbre control, environmental acoustic consistency with text prompts, and fidelity in preserving source linguistic content.
\subsubsection{Subjective Evaluation}
We benchmark TES-VC against VoiceLDM to evaluate timbre accuracy, environmental consistency, and audio naturalness. Our test set comprises 200 speaker descriptions generated by GPT-4o, each specifying gender, age (e.g., boy, young man/man/old male), and randomized timbral attributes (e.g., rough, deep, hoarse). These prompts are paired with environmental descriptions randomly sampled from the AudioCaps-filter subset. Source speech is drawn from the libriTTS-R-test-clean corpus. For fair comparison, we concatenate environmental and timbre descriptions as VoiceLDM’s input prompts while using ground-truth transcripts as content prompts. All evaluations are conducted through subjective listening tests to eliminate biases from linguistic content variations.

Our subjective assessment comprises both quantitative accuracy metrics and qualitative Mean Opinion Score (MOS) evaluations: for quantitative analysis, human annotators labeled the perceived age and gender of generated samples, based on which we calculated age/gender recognition accuracy; for qualitative assessment, we adopt three MOS criteria – Speaker Consistency (S-MOS), Environmental Faithfulness (E-MOS) and Realness (R-MOS) – where fifteen participants evaluated randomly selected samples using a 5-point Likert scale.

As shown in Table 1, our method demonstrates significant advantages over VoiceLDM across all subjective evaluation metrics. Notably, VoiceLDM exhibits substantially weaker performance in quantitative age/gender accuracy and S-MOS, primarily due to CLAP's tendency to discard fine-grained details when processing prompts containing both vocal and environmental descriptions, severely limiting its timbre control capability. While CLAP's feature suppression also mildly impacts environmental fidelity (lower E-MOS), this effect is less prominent compared to timbre-related degradation, though it occasionally leads to missing ambient sounds. In contrast, TES-VC achieves decoupled control over environment and speaker characteristics through its specialized architecture, outperforming VoiceLDM's single-CLAP framework in precision. Furthermore, the proposed method attains significantly higher R-MOS scores, mainly because VoiceLDM may produce unreasonable volume, also proving that synthetic training data can yield realistic audio outputs comparable to genuine recordings.
\begin{figure} 
    \centering
    \makebox[\columnwidth]{\includegraphics[scale=0.4]{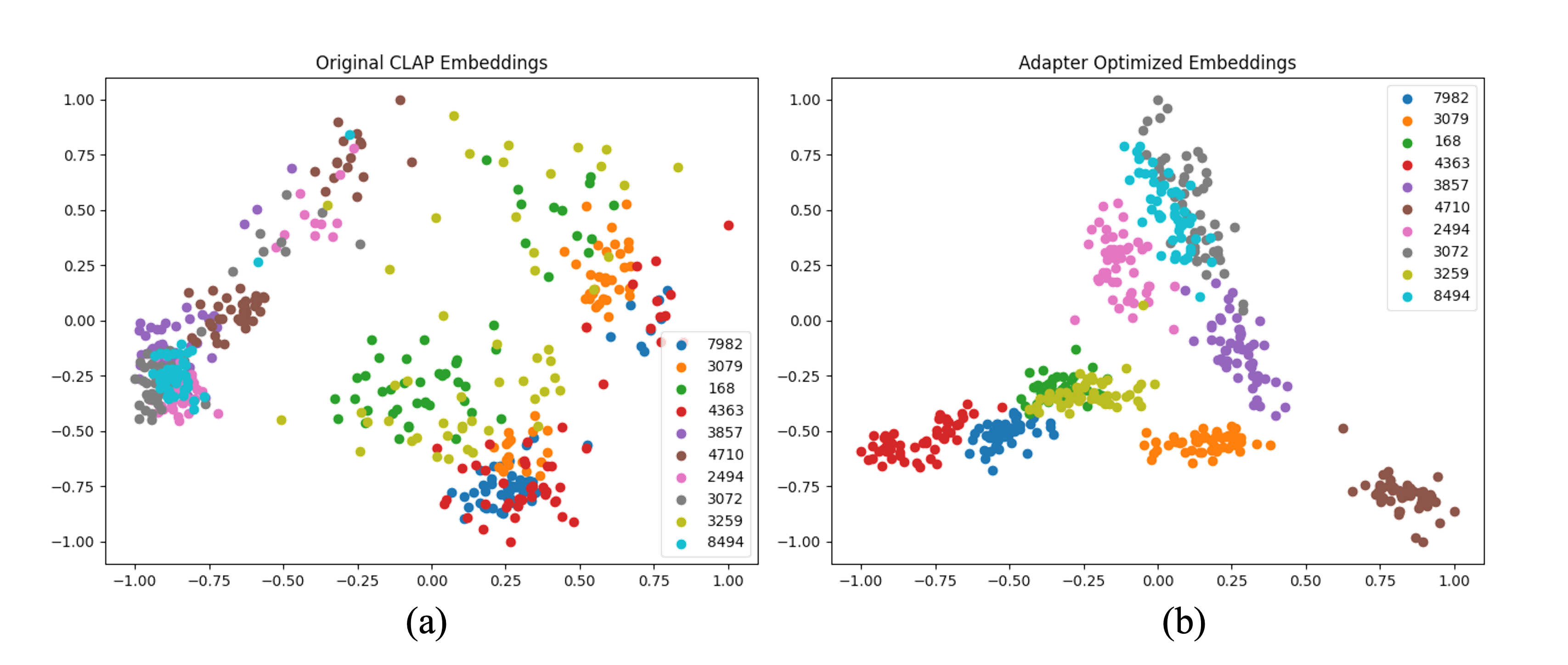} }
    \caption{The visualization analysis via PCA. Subfigure (a) shows the distribution of original CLAP embeddings. Subfigure (b) shows the distribution of the optimized embeddings. The number in labels refers to the speaker IDs in Librispeech dataset.}
    \label{fig:example}
\end{figure}
\begin{table}
  \caption{Experimental results of objective evaluation.}
  \label{tab:example}
  \centering
    \begin{tabular}{c c c c c}
    \toprule
    \textbf{Systems} & \textbf{WER$\downarrow$} & \textbf{CER$\downarrow$} & \textbf{KL$\downarrow$} & \textbf{CLAP$\uparrow$} \\
    \midrule
    Ground Truth        & $8.96\%$  & $2.82\%$     & $-$     & $0.259$   \\
    \midrule
    VoiceLDM\cite{lee2024voiceldm}        & $-$  & $-$     & $1.736$     & $0.176$   \\
    FreeVC\cite{li2023freevc}        & $\bm{11.95\%}$  & $5.92\%$     & $-$     & $-$   \\
    \textbf{TES-VC}            & $12.09\%$  & $\bm{5.55\%}$     &  $\bm{1.704}$      & $\bm{0.183}$   \\
    \bottomrule
    \end{tabular}
\end{table}
\subsubsection{Objective Evaluation}
To comprehensively evaluate the proposed method, we conducted two-fold objective assessments. For content preservation, we compared our method with FreeVC \cite{li2023freevc} on a standard voice conversion task using 200 audio samples from the LibriTTS-R-Test dataset, with reference audio from 8 unseen speakers and environmental control input set to ``clean speech". Word Error Rate (WER) and Character Error Rate (CER) were measured via whisper-large-v2 \cite{radford2023robust}. To assess environmental sound generation capability, we disabled content inputs of both VoiceLDM and our method, retaining only environmental text prompts. Tests on the AC-filter test set utilized the CLAP score and Kullback-Leibler (KL) divergence as metrics.

The objective results in Table 2 demonstrate that our method achieves comparable WER and CER to FreeVC. This indicates that despite incorporating environmental control conditions and introducing substantial non-speech data during training, our approach effectively preserves the content integrity of source speech. Furthermore, our method slightly outperforms VoiceLDM in CLAP scores and KL metrics, which can be attributed to our synthetic data training strategy. These results also validate the method's robust capability in generating environmental sound effects.
\subsubsection{Ablation Study}
To validate the efficacy of our proposed RBTC module, we conduct an ablation study (Table 1, last row). Removing the CLAP-Timbre Adapter results in reduced age/gender control accuracy and a statistically significant drop in S-MOS (Speaker Consistency). This degradation stems from the inherent limitations of raw CLAP embeddings: while they partially encode speaker characteristics, these features are often entangled with unrelated factors (e.g., linguistic content or environmental acoustics), leading to ambiguous timbre representations. Our adapter enhances speaker-specific discriminability in CLAP embeddings, yielding more precise age/gender alignment and higher S-MOS.
\subsubsection{Visualization Analysis}
We further visualize the distributional differences between raw and adapted CLAP embeddings to quantify speaker-discriminative improvements. Using 50 audio clips per speaker from 10 randomly selected LibriSpeech speakers, we project both embedding types into 2D space via principal component analysis (PCA). As shown in Figure 2, adapted embeddings exhibit distinct speaker-specific clustering patterns compared to the diffusely distributed raw CLAP features, empirically confirming that our adaptation strengthens the correlation between CLAP representations and speaker identity.

\section{Conclusion}
This paper presents TES-VC, a novel voice conversion framework enabling text-driven independent control over speaker timbre and acoustic environments. Our systematic data construction methodology facilitates disentangled learning of content preservation, environmental acoustics, and speaker characteristics. The proposed Retrieval-Based Timbre Control (RBTC) module achieves diverse speaker manipulation without requiring text-speech paired training data. Subjective and objective evaluations validate TES-VC's capability to precisely control vocal characteristics and acoustic environments while maintaining source speech content integrity, demonstrating the efficacy of the proposed methodology.

\section{Acknowledgements}
This work is supported by National Natural Science Foundation of China (62076144), National Social Science Foundation of China (13\&ZD189) and Shenzhen Science and Technology Program (JCYJ20220818101014030).

\bibliographystyle{IEEEtran}
\bibliography{mybib}

\end{document}